\documentclass[twocolumn,aps,prl]{revtex4}
\usepackage{epsfig,amsmath,graphicx,gensymb,textcomp}

\begin{document}

\title{Absolute Measurement of Quantum-Limited Interferometric Displacements}

\author{Val\'{e}rian Thiel, Pu Jian, Claude Fabre, Nicolas Treps, Jonathan Roslund}

\affiliation{\mbox{Laboratoire Kastler Brossel, UPMC-Sorbonne Universit\'es, CNRS, ENS-PSL Research University, Coll\`ege de France} 4 place Jussieu, 75252 Paris, France\\
Corresponding author: jonathan.roslund@lkb.upmc.fr}

\date{\today}

\begin{abstract}

A methodology is introduced that enables an absolute, quantum-limited measurement of subwavelength interferometric displacements. The technique utilizes a high-frequency optical path modulation within an interferometer operated in a homodyne configuration. All of the information necessary to fully characterize the resultant path displacement is contained within the relative strengths of the various harmonics of the phase modulation. The method, which is straightforward and readily implementable, allows a direct measurement of the theoretical Cram\'{e}r-Rao limit of detection without any assumptions on the nature of the light source.

\end{abstract}

\maketitle

\section{Introduction}

The measurement of distances smaller than the wavelength of light is fundamental to precision metrology and is routinely accomplished with optical interferometry \cite{abramovici1992ligo}. The smallest distance variations that may be registered by an interferometer quite naturally define the device's detection limit. For the situation in which the only noise source contributing to a measurement arises from the quantum nature of light (i.e., shot noise fluctuations), this detection limit is well-known \cite{caves1981quantum} and scales as $d_{\textrm{min}} \sim \lambda / \sqrt{N}$, where $\lambda$ is the wavelength of light and $N$ is the number of photons in the light field.
For modest power levels ($\sim 1 \; \textrm{mW}$), wavelengths on the order of $\sim 1$ \textmu \textrm{m}, and integration times of approximately $1~\textrm{ms}$, the theoretical limit is a fraction of an optical fringe (e.g., $ d_{\textrm{min}}  < 10^{-12}~\textrm{m}$), which renders it considerably challenging to measure in a direct manner.
%
%
As a result, such distances are typically quantified indirectly by measuring the parameters that characterize the quantum limit, such as the optical power \cite{treps2003quantum,verlot2009scheme}. An accurate determination of these field quantities often proves cumbersome and is susceptible to significant error. For instance, measurement of the photon number $N$ demands a precisely calibrated power meter, knowledge of the diode efficiency, assurance that all of the photons are reaching the diode, the fringe contrast of mode-matching, etc. Thus, it is of interest to be able to measure such small displacements in a direct manner. Toward this end, we propose and implement a method that enables an absolute calibration of subwavelength interferometric displacements and is limited only by the intrinsic quantum mechanical fluctuations of the light field. The technique utilizes a high-frequency modulation of the optical path in one arm of an interferometer. A Fourier analysis of the modulation harmonics in the interferometer signal enables the observation of optical path displacements on the order of $d_{\textrm{min}} \sim 10^{-15} \, \textrm{m}$. With the ability to directly measure such subwavelength path variations, the experimentally observed displacements are compared to and shown to coincide with the theoretical quantum limit \cite{caves1981quantum,lamine2008quantum}.

\section{Methodology}
\label{sec:methods}

The present work employs an interferometer that operates in a homodyne configuration. In particular, a single laser field is divided with a $90:10$ beamsplitter, and the output ports constitute the two arms of the interferometer. One arm carries a strong local oscillator (LO) field $E_{LO}$ (the $ 90\%$ port of the beamsplitter) while the other arm contains a weaker signal field $E_{\textrm{sig}}$. These two fields are taken to be complex and are decomposed in terms of their spectral amplitudes $A$ and phases $\phi$ in a form described as $E_{LO} = \mathcal{E}_{0} A_{LO} f_{LO}(\vec{r}) \exp[ - i \omega_{0} t ]$ and $E_{\textrm{sig}} = \mathcal{E}_{0} A_{\textrm{sig}} f_{\textrm{sig}}(\vec{r}) \exp[ i ( \phi_{\textrm{rel}} - \omega_{0} t) ]$, where $\mathcal{E}_{0}$ is the field amplitude for a single photon\footnote{The field amplitude $\mathcal{E}_{0} = \sqrt{ \hbar \omega_{0} / 2 \epsilon_{0} V }$ is specified by quantifying the electric field in a box of volume $V$ \cite{bachor2004guide}.}, $f(\vec{r})$ is the spatial profile of the beam, $\omega_{0}$ is the optical carrier frequency, and $\phi_{\textrm{rel}}$ is the relative phase difference between the two interferometer arms, i.e., $\phi_{\textrm{rel}} = \phi_{LO} - \phi_{\textrm{sig}}$.
The spectral amplitudes $A$ determine the average number of photons carried by the individual fields through the relation $A = \sqrt{N}$. Both of these fields travel through the interferometer and are recombined on a $50:50$ beamsplitter in which each output port is independently detected with a photodiode. The difference of photocurrents between the two diodes yields the homodyne signal $ S_{\textrm{HD}} = R \cdot \int \left( E_{LO} E_{\textrm{sig}}^{*}  + c.c. \right) \textrm{d}\vec{r}$, where $R = \Phi_{q} \, e / \hbar \omega_{0} $ specifies the diode response with  $\Phi_{q}$ representing the quantum efficiency of the diode and $e$ is the electron charge \cite{bachor2004guide}. Given the field representations considered above, the homodyne signal is rewritten as:
\begin{equation} \label{eq:homodyne}
S_{\textrm{HD}} = 2 \, \mathcal{E}_{0}^{2} \, \eta \cdot A_{LO} A_{\textrm{sig}} \cos \left[ \phi_{\textrm{rel}} \right],
\end{equation}
where $\eta = R \cdot \gamma$ and $\gamma = \int \textrm{d}\vec{r} \, f_{\textrm{sig}}^{\ast} \, f_{LO}$ is the overlap between the two fields\footnote{For the situation in which broadband light is utilized, the envelopes $f$ are also taken to be spectrally dependent and $\gamma$ denotes both the spectral and spatial mode overlap.} that defines the contrast of the interference fringes.

A mirror in one arm of the interferometer is mounted upon a piezoelectric actuator, which allows well-controlled path displacements $d(V)$ that are to be calibrated. Sinusoidal movement of the mirror-piezo system generates a corresponding modulation of the optical path between the two interferometer arms that is described by:

\begin{equation}
\phi_{\textrm{mod}} = k \cdot n \cdot d(V) \cdot \sin \left[ \Omega t \right],
\end{equation}
where $k = 2 \pi / \lambda$ is the field wavenumber, $n$ is the refractive index for the material in which the modulation occurs, $d(V)$ is the voltage-controlled displacement that is to be determined, and $\Omega$ is the modulation frequency. It is worth mentioning that the implicit assumption of a linear piezo response to the applied voltage is later confirmed (e.g., see Fig.~\ref{fig-piezo-scan}a).

The composite relative phase between the two interferometer arms is then written as $\phi_{\textrm{rel}} = \phi_{0} + \phi_{\textrm{mod}}$, where $\phi_{0}$ represents the mean phase offset combined with any noise fluctuations of the relative phase (i.e., $\phi_{0} = \langle \phi_{0} \rangle + \delta \phi_{0}$). Accordingly, the homodyne signal of the interferometer becomes $S_{\textrm{HD}} = 2 \, \mathcal{E}_{0}^{2} \, \eta \, A_{LO} A_{\textrm{sig}} \cos \left[ \phi_{0} + \phi_{\textrm{mod}} \right]$. Since the modulation amplitude $d$ is taken to be much smaller than the optical wavelength (i.e., $k \cdot d \ll 1$), the trigonometric function of the homodyne signal is expanded in a basis of the modulation harmonics to yield

\begin{eqnarray}
S_{\textrm{HD}} &\simeq& 2 \, \mathcal{E}_{0}^{2} \, \eta \cdot A_{LO} A_{\textrm{sig}} \left[ J_{0} (k \, n \, d) \cos \phi_{0} \right. \\ \nonumber
&-& 2 \, J_{1} (k \, n \, d) \sin \phi_{0} \sin \left[ \Omega t \right] \\ \nonumber
&+& \left. 2 \,J_{2} (k \, n \, d) \cos \phi_{0} \cos \left[ 2 \, \Omega t \right] + \ldots \right].
\end{eqnarray}
The terms $J_{n}$ are Bessel functions of the first kind, and the displacement $d$ is implicitly taken to be a function of the voltage applied to the piezo actuator. The signal at the fundamental frequency $\Omega$ is maximized by setting the mean phase offset to $\langle \phi_{o} \rangle = \pi / 2$, which corresponds to an interrogation of the signal field's phase quadrature.
%
%
Conversely, the second harmonic of the modulation appears as a modulation of amplitude, and is thus detected when the LO is in phase with the signal field (i.e., $ \langle \phi_{0} \rangle = 0$).


The detected signal is also subject to noise variations $\delta S_{\textrm{HD}}$ that arise from fluctuations of the optical field.
In line with a homodyne detection scheme, the LO field amplitude $A_{LO}$ is significantly larger than the signal amplitude $A_{\textrm{sig}}$, i.e., $A_{LO} \gg A_{\textrm{sig}}$. Under this assumption, the variance of the photocurrent signal is calculated from Eq.~\ref{eq:homodyne} in the absence of a modulation (i.e. $\phi_{\textrm{mod}} = 0$) to be

\begin{equation} \label{eq:hd-variance}
\langle \delta S_{\textrm{HD}}^{2} \rangle = 4 \, \mathcal{E}_{0}^{4} \, \eta^{2} \cdot A_{LO}^{2} \left[ \langle \delta A_{\textrm{sig}}^{2} \rangle \cos^{2} \phi_{0} + A_{\textrm{sig}}^{2} \langle \delta \phi^{2}_{0} \rangle \sin^{2} \phi_{0} \right],
\end{equation}
where $\delta A_{\textrm{sig}}$ and $\delta \phi_{0}$ are the fluctuations of the signal field amplitude and relative phase offset, respectively \cite{bachor2004guide}. If the modulation is performed at a frequency for which the optical source does not possess any noise other than that arising from vacuum fluctuations (i.e., shot noise limited), the two quadrature variances are readily evaluated to be $\langle \delta A_{\textrm{sig}}^{2} \rangle = A_{\textrm{sig}}^{2} \langle \delta \phi_{0}^{2} \rangle = 1 / 4$\footnote{The amplitude and phase of the signal may be written in terms of the field quadratures $X$ and $P$ as $X = 2 A \cos \phi$ and $P = 2 A \sin \phi$. For quantum-limited noise fluctuations, the fluctuations of these quadrature operators are equal and possess the value $\langle \delta X ^{2} \rangle = \langle \delta P ^{2} \rangle = 1.0$ \cite{bachor2004guide}. As such, the quantum limit for the amplitude and phase fluctuations is $\langle \delta A ^{2} \rangle = A^{2} \langle \delta \phi ^{2} \rangle = 1/4$.}. It is also important to stress that operation at the shot noise limit allows the removal of quadrature correlations from Eq.~\ref{eq:hd-variance}, i.e., $\langle \delta A_{\textrm{sig}} \delta \phi \rangle = 0$.

The noise of the measurement is therefore quadrature-independent and given by $\Delta S_{\textrm{HD}} = \mathcal{E}_{0}^{2} \, \eta \cdot A_{LO}$. The signal-to-noise $\Sigma$ of the homodyne signal is finally written in the form:

\begin{eqnarray} \label{eq:hd-signal-to-noise}
\Sigma &\simeq& 2 \, A_{\textrm{sig}} \left[ J_{0} (k \, n \, d) \cos \phi_{0} \right. \\ \nonumber
&-& 2 \, J_{1} (k \, n \, d) \sin \phi_{0} \sin \left[ \Omega t \right] \\ \nonumber
&+& \left. 2 \,J_{2} (k \, n \, d) \cos \phi_{0} \cos \left[ 2 \, \Omega t \right] + \ldots \right].
\end{eqnarray}
The intensity of individual harmonics of $\Sigma$ clearly depends upon the energy contained within the signal field. The signal-to-noise of the first and second-harmonics of the homodyne signal are represented as $\Sigma^{(1)} = - 4 \, A_{\textrm{sig}} \, J_{1} (k \, n \, d) $ and $\Sigma^{(2)} = 4 \, A_{\textrm{sig}} \,J_{2} (k \, n \, d) $, respectively.
For small displacements (i.e., $k \cdot d \ll 1$), the signal-to-noise at the fundamental frequency may be written as $ \left| \Sigma^{(1)} \right| = 2 \, A_{\textrm{sig}} \, \delta \phi_{\textrm{mod}}$, where $\delta \phi_{\textrm{mod}} = k \, n \, d$. A signal-to-noise of $\left| \Sigma^{(1)} \right| = 1.0$ defines the detection limit for the instrument, which implies that the smallest observable phase modulation is $\delta \phi_{\textrm{mod}}^{\textrm{HD}} = 1 / \left( 2 \sqrt{N} \right)$, where $A_{\textrm{sig}} = \sqrt{N}$ \footnote{It is worth noting that the homodyne-based limit is smaller than the standard interferometer limit by a factor of two \cite{caves1981quantum}. This differences arises from the fact that the homodyne implementation places the $N$ signal photons into a single arm of the interferometer while the standard interferometer distributes them equally between both arms}. It may be shown that this limit coincides with the Cram\'{e}r-Rao bound (i.e., the standard quantum limit) for phase estimation \cite{lamine2008quantum}, which indicates that the measurement of phase displacements with homodyne-based interferometry yields the standard quantum limit, i.e., $ \delta \phi_{\textrm{mod}}^{\textrm{HD}} = \delta \phi_{\textrm{mod}}^{\textrm{SQL}} $ \cite{pinel2012ultimate}.

The ratio of $ \Sigma^{(1)} $ and $ \Sigma^{(2)} $ is then examined, which is given by
%
%
\begin{equation} \label{eq:ratio}
\frac{\Sigma^{(2)}}{\Sigma^{(1)}} = \left| \frac{J_{2} (k \, n \, d)}{J_{1} (k \, n \, d)} \right| \simeq \frac{k \, n \, d}{4}.
\end{equation}
This ratio is independent of the signal field strength and varies linearly with the optical displacement. By varying the mean phase offset between $ \langle \phi_{0} \rangle = \pi / 2$ and $\langle \phi_{0} \rangle = 0$, the signal level of these individual harmonics may be measured, and their ratio provides an intrinsic means for calibrating the optical displacement.

\section{Experimental Details}

\begin{figure}[htbp]
\centering
\fbox{\includegraphics[width=65mm]{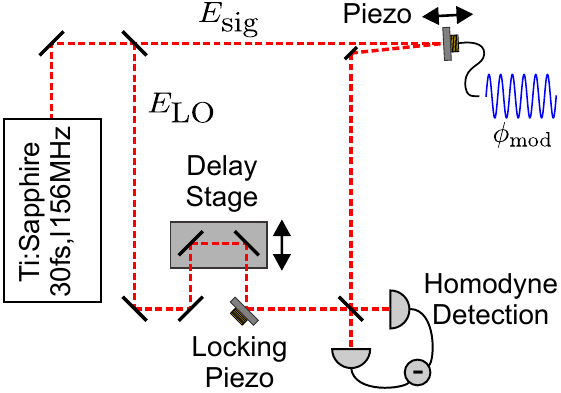}}
\caption{Experimental layout for calibrating the subwavelength displacements of optical path.
The source is split, and the signal portion of the beam impinges on a mirror that is mounted upon a piezoelectric actuator. This piezo element is driven with a high frequency sinusoidal voltage source. The local oscillator (LO) arm of the interferometer is delayed appropriately (not necessary for continuous wave applications), and the two beams are recombined on a $50:50$ beamsplitter. The outputs of this beamsplitter are analyzed with homodyne detection. An additional piezo-controlled mirror in the LO arm allows for controlling the mean relative phase $\langle \phi_{0} \rangle$ between the two fields.}
\label{fig-exp-setup}
\end{figure}

The laser source is a titanium:sapphire modelocked system delivering $\sim 30 \textrm{fs}$ pulses centered at 795nm with a repetition rate of $156 \textrm{MHz}$. It is important to note, however, that an ultrafast system is not required to implement this calibration methodology, which is equally applicable for continuous wave sources. The pulses are delivered to the entrance of a Mach-Zehnder interferometer, and the powers are adjusted to have $\sim 115 \mu \textrm{W}$ in the signal field and $\sim 1.5 \textrm{mW}$ for the LO. As seen in Fig.~\ref{fig-exp-setup}, a mirror mounted upon a piezo element (Physik Instrumente, P-010.00H) in the signal arm of the interferometer delivers the high-frequency phase modulation $\phi_{\textrm{mod}}$. The piezoelectric actuator is driven at 2MHz with a voltage that ranges between $0$ and $10$ Volts. The signal beam is incident on this piezo-mounted element at an angle of $< 1 \degree$ in order to generate a purely longitudinal phase variation. A second piezo element is located in the LO arm of the interferometer and allows for controlling the mean optical phase offset $ \langle \phi_{0} \rangle$.

The two beams are recombined on a $50:50$ beamsplitter and subsequently detected with a pair of balanced and amplified silicon photodiodes (Thorlabs, PDA36A). The high- and low-frequency components of each diode's photocurrents are separated at $\sim 10 \textrm{kHz}$. The difference of the two high-frequency (HF) photocurrents is analyzed with a spectrum analyzer. Traces are collected in a zero-span mode centered around the modulation frequency of $\sim 2 \textrm{MHz}$ with resolution and video bandwidths of $50 \textrm{Hz}$ and $5 \textrm{Hz}$, respectively. Individual traces from the spectrum analyzer are transferred to a computer for processing.

In order to observe the evolution of the first harmonic $\Sigma^{(1)}$, the relative phase between the two beams is locked to $ \langle \phi_{0} \rangle = \pi / 2$ by utilizing the difference of the two low-frequency (LF) photocurrents as an error signal. Conversely, the growth of the second harmonic $\Sigma^{(2)}$ is examined under the same voltage range by locking the mean relative phase to $ \langle \phi_{0} \rangle = 0 $. This is accomplished by demodulating a portion of the HF difference signal with an electronic mixer and a RF local oscillator of the same frequency that drives the piezo movement. This demodulated homodyne signal provides an error signal.

\section{Results}

\subsection{Calibration}

\begin{figure}[htbp!]
\centering
\includegraphics[width=\linewidth]{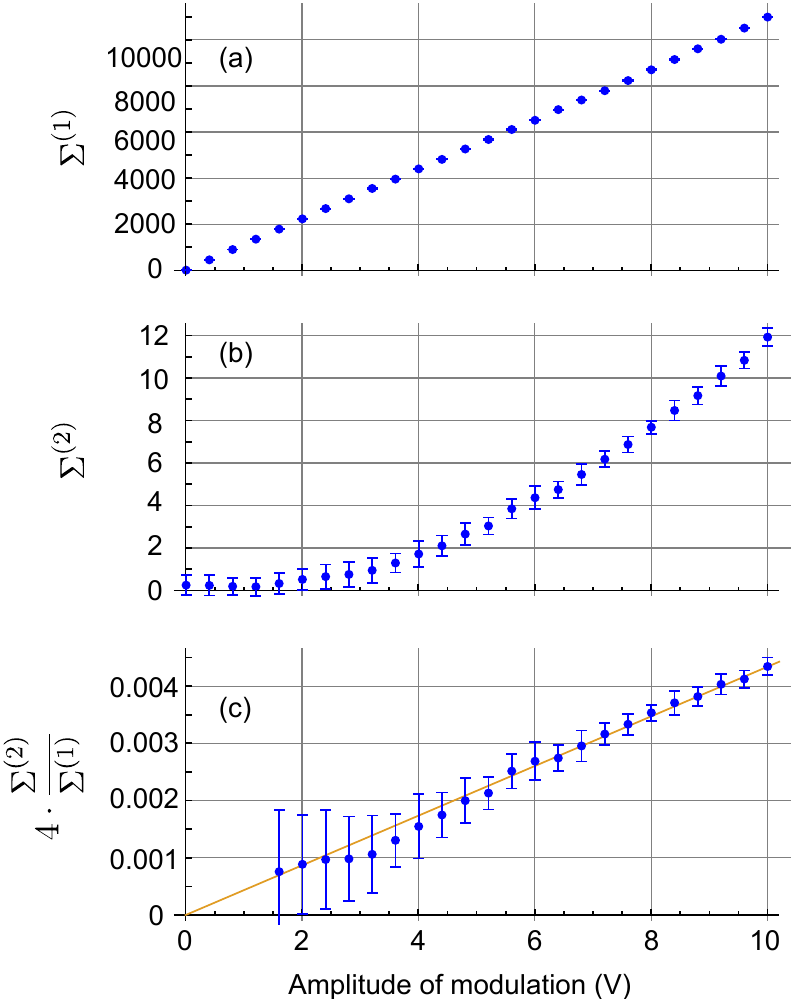}
\caption{Phase modulation harmonics observed in the homodyne signal as a function of the voltage applied to the piezoelectric actuator. The fundamental harmonic (a) is observed in the phase-quadrature of the signal field and grows linearly with the optical displacement. The second harmonic (b) is detected in the amplitude quadrature of $E_{\textrm{sig}}$ and depends quadratically on the piezo movement. The ratio of the two harmonics (c) is linearly dependent upon the phase retardation induced by the piezo. The retrieved piezo sensitivity is $0.55 \pm 0.01 \textrm{\AA} / \textrm{Volt}$.}
\label{fig-piezo-scan}
\end{figure}

The evolution of the amplitudes for the fundamental and second harmonic are shown in Fig.~\ref{fig-piezo-scan}a and b, respectively, as a function of the voltage supplied to the piezo. The fundamental harmonic exhibits a high signal-to-noise $\Sigma^{(1)}$ and grows linearly for small displacements of the path (i.e., $ J_{1}[ \phi ] \simeq \phi / 2$). The observation of a linear growth for $\Sigma^{(1)}$ also confirms that the piezo response itself is linear for the investigated voltage range. The signal-to-noise of the second harmonic $\Sigma^{(2)}$ is markedly smaller and grows quadratically with the path displacement (i.e., $J_{2}[\phi] \simeq \phi^{2} / 8$).
As seen from Eq.~\ref{eq:hd-signal-to-noise}, each of these signals is dependent upon the number of photons contained within the signal field $E_{\textrm{sig}}$.

The ratio of the two harmonics, however, is independent of the signal field strength and is displayed in Fig.~\ref{fig-piezo-scan}c. As expected, the ratio is linear with respect to the voltage applied to the piezo, and a linear fit of this data according to Eq.~\ref{eq:ratio} yields an optical displacement of $0.55 \pm 0.01 \textrm{\AA} / \textrm{Volt}$ \footnote{The calibration procedure is performed in laboratory air. As such, the index of refraction is taken to be $n = 1.0$.}.


\subsection{Sensitivity Measurement}

\begin{figure}[htbp]
\centering
\includegraphics[width=75mm]{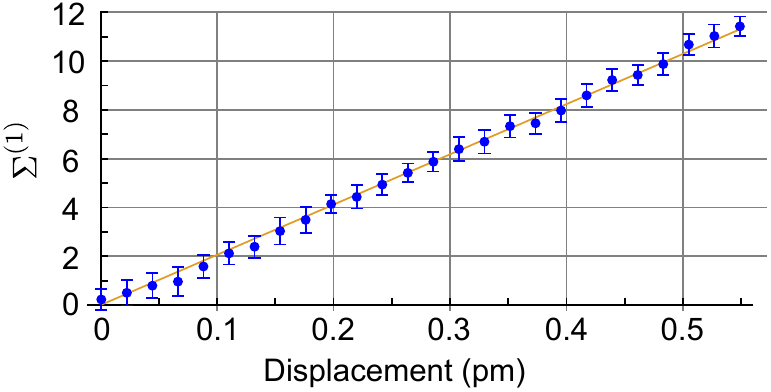}
\caption{Time-averaged displacement sensitivity of the mirror mounted upon the calibrated piezo element. For an integration of ~9 milliseconds ($\textrm{RBW} = 50\textrm{Hz}$), which amounts to $N = 2.7 \pm 0.2 \cdot 10^{12}$ photons, the time-averaged detection limit (i.e., $\Sigma^{(1)} = 1.0$) is $d_{\textrm{min}} = 5.0 \pm 0.1 \cdot 10^{-14} \, \textrm{m}$. Following a normalization by this integration time, the smallest detectable path change is $d_{ \textrm{min} / \sqrt{\textrm{Hz}} } = 7.0 \pm 0.1 \cdot 10^{-15} \, \textrm{m} / \sqrt{\textrm{Hz}}$.}
\label{fig-piezo-sensitivity}
\end{figure}

Given this calibration, it becomes possible to directly measure the phase sensitivity of the interferometer. As before, the interferometer is locked on the phase quadrature of the signal field, and the signal-to-noise ratio of the fundamental harmonic is monitored as the voltage is scanned over a reduced range. The observed sensitivity is displayed in Fig.~\ref{fig-piezo-sensitivity}. The minimum detectable amplitude of modulation $d_{\textrm{min}}$ occurs for a signal that is equal in magnitude to the background quantum noise, i.e., $\Sigma^{(1)} = 1.0$. From Fig.~\ref{fig-piezo-sensitivity}, it may be observed that the time-averaged sensitivity is $d_{\textrm{min}} = 5.0 \pm 0.1 \cdot 10^{-14} \, \textrm{m}$ (equivalent to a phase precision of $0.40~\mu \textrm{rad}$).
%
%
The sensitivity deduced from Fig.~\ref{fig-piezo-sensitivity} results from collecting photons during a time interval corresponding to a resolution bandwidth of 50Hz. Upon normalizing the recovered sensitivity by the noise bandwidth, the minimum detectable displacement becomes $d_{ \textrm{min} / \sqrt{\textrm{Hz}} } = 7.0 \pm 0.1 \cdot 10^{-15} \, \textrm{m} / \sqrt{\textrm{Hz}}$ (equivalent to a phase precision of $5.6 \cdot 10^{-8}~\textrm{rad} / \sqrt{\textrm{Hz}}$).


From the Cram\'{e}r-Rao bound discussed above, the displacement amplitude corresponding to the standard quantum limit is written as  $d_{\textrm{min}}^{\textrm{SQL}} = \lambda_{0} / \left( 4 \pi \, \sqrt{N} \right)$. The number of signal photons $N$ is determined from both the optical power of the signal beam $P$ and the integration time of the detection system $\Delta t$ (i.e., $N = P \cdot \Delta t / \hbar \omega_{0} $).
The swept spectrum analyzer utilizes a nearly Gaussian spectral filter for the noise measurements, which allows the integration time $\Delta t$ to be written in terms of the $-3 \textrm{dB}$ resolution bandwidth $\Delta \nu_{\textrm{RBW}}$ as $\Delta t \simeq 0.44 / \Delta \nu_{\textrm{RBW}}$.
The calculation of $N$ with the employed experimental parameters reveals that the theoretical standard quantum limit is $d_{\textrm{min}/\sqrt{\textrm{Hz}}}^{\textrm{SQL}} \simeq 5.4 \pm 0.2 \cdot 10^{-15} \, \textrm{m} / \sqrt{\textrm{Hz}}$. Thus, the experimentally recovered sensitivity is in fair agreement with that predicted by the standard quantum limit ($d_{\textrm{min}/\sqrt{\textrm{Hz}}} / d_{\textrm{min}/\sqrt{\textrm{Hz}}}^{\textrm{SQL}} \simeq 1.3$).

It is important to stress that the present observation of subwavelength displacements ($\sim 10^{-12} \, \textrm{m}$) corresponds to the time-averaged oscillation amplitude of the mirror under question. As such, these average positions are determined by integrating the photon flux for a comparably long timescale (i.e., $\sim \textrm{ms}$). Thus, fluctuations that occur on a timescale shorter than the integration time are averaged away and not observed.

\section{Discussion}

Importantly, this calibration strategy does not necessitate an initial observation of optical fringes, which allows for fitting the observed response to a sinusoidal function. In such a case, several orders of magnitude separate the displacements of interest ($\lesssim 10^{-12}$) from those necessary to observe fringes ($\sim 10^{-6}$), and it must be assumed that both the piezo actuator and driving electronics operate linearly over this entire range. The determination of path changes from the relative strengths of various modulation harmonics does not necessitate any assumptions.

Additionally, the role of classical noise in an interferometer is often mitigated by transferring slowly-varying motions of interest to the sidebands of a high-frequency modulation \cite{stevenson1993quantum,taylor2013biological}. With a reliable calibration, these absolute movements may be directly determined.

It is also worth mentioning that this calibration strategy could conceivably be extended to modulation frequencies that are not quantum noise limited (i.e., classical noise fluctuations occur on the timescale of the modulation).
If an accurate accounting of the quadrature-dependent classical noises are made at the relevant detection frequencies and intra-quadrature noise correlations may be neglected (i.e., $\langle \delta A_{\textrm{sig}} \delta \phi \rangle = 0$), the ratio of the two signal-to-noises becomes $\Sigma^{(2)} / \Sigma^{(1)} \simeq \left[ \left( k \, n \, d\right) / 4 \right] \cdot \left( A_{\textrm{sig}} \, \Delta \phi_{\textrm{sig}}^{(1)} / \Delta A_{\textrm{sig}}^{(2)} \right)$, where $A_{\textrm{sig}} \Delta \phi_{\textrm{sig}}^{(1)}$ is the phase noise relative to the shot noise limit at the fundamental modulation frequency. Conversely, $\Delta A_{\textrm{sig}}^{(2)}$ is the amplitude noise relative to the shot noise limit at the second harmonic of the modulation frequency.

Finally, although the present application examined the calibration of a mirror-piezo system placed in an interferometer, it has equally well been applied for the determination of phase displacements induced by an electro-optic modulator (EOM).

\section{Conclusions}

In conclusion, this work presents a shot noise limited methodology that allows an absolute measurement of time-averaged subwavelength motions in an interferometer. These experimentally determined displacements allow measurement of the theoretical Cram\'{e}r-Rao bound.
The strategy is readily-implementable using standard laboratory equipment and does not require any assumptions regarding theoretical detection limits. Consequently, this approach may find utility in a range of domains, including optical metrology, optomechanics, and quantum optics.

\bigskip

The work was supported by the European Research Council starting grant Frecquam. J.R. acknowledges support from the European Commission through Marie Curie Actions QOCO.



\bibliographystyle{apsrev}
\bibliography{piezo-calib-BIB}

\end{document}